\documentclass[12pt,epsf]{article}
\usepackage{graphicx, amsmath}

\begin{document}
\sloppy
\begin{flushright}{SIT-HEP/TM-54}
\end{flushright}
\vskip 1.5 truecm
\centerline{\large{\bf Evolution of curvature perturbation in
generalized gravity theories}}
\vskip .75 truecm
\centerline{\bf Tomohiro Matsuda\footnote{matsuda@sit.ac.jp}}
\vskip .4 truecm
\centerline {\it Laboratory of Physics, Saitama Institute of Technology,}
\centerline {\it Fusaiji, Okabe-machi, Saitama 369-0293, 
Japan}
\vskip 1. truecm

\makeatletter
\@addtoreset{equation}{section}
\def\theequation{\thesection.\arabic{equation}}
\makeatother
\vskip 1. truecm

\begin{abstract}
\hspace*{\parindent}
Using the cosmological perturbation theory in terms of the $\delta N$
 formalism, we find the simple formulation of the evolution of the
 curvature perturbation in generalized gravity theories.
Compared with the standard gravity theory, a crucial difference 
appears in the end-boundary of the inflationary stage, which is due to
 the non-ideal form of the energy momentum tensor that depends
 explicitly on the curvature scalar.
Recent study shows that ultraviolet-complete quantum theory of
 gravity may be approximated by using a
 generalized gravity action.
Our paper may give an important  step in understanding the evolution of
 the 
 curvature perturbation during inflation, where the energy momentum
 tensor may not be given by the ideal form due to the corrections from
 the fundamental theory.
\end{abstract}

\newpage
\section{Introduction}
The inflationary Universe scenario is the most successful model for 
explaining the large scale structure of the Universe.
The spectrum of the cosmological perturbations created during
inflation is expected to be scale-invariant and Gaussian.
An inflationary Universe is consistent with current observations of the
temperature anisotropy of the cosmic microwave background (CMB), except
 for some anomalies in the spectrum.
These anomalies include a small departure from exact scale-invariance
and a certain non-Gaussian character \cite{Bartolo-text, NG-obs},
both of which can help reveal the underlying gravitational
 and field theoretical model as well as the dynamics of the fields during
 inflation. 
There are many models of the mechanism for generating curvature
perturbations, which can be categorized either in
terms of (A) when and (B) how they are created.
Focusing on the question of when perturbations are generated, 
curvature perturbations can be generated (A-1) at the horizon exit
 \cite{EU-book}, (A-2) by the evolution during inflation
 \cite{Modulated-matsuda, roulette-inflation},
(A-3) at the inhomogeneous end of inflation \cite{End-Modulated,
End-multi-mat}, (A-4) just after inflation during
preheating/reheating\cite{IH-PR, IH-R, Preheat-ng},    
 or (A-5) late after inflation by curvatons \cite{curvaton-paper,
 matsuda_curvaton}
or by an inhomogeneous phase transition\cite{IH-pt}. 
In this paper, we consider the curvature perturbations in generalized
 gravity theories for (A-1), (A-2) and (A-3): 
the generation at the horizon crossing, evolution during inflation
and the correction at the inhomogeneous end of inflation.
Focusing on the mechanism of generating curvature perturbations,
curvature perturbations can be created by 
(B-1) perturbations at the boundary between different scaling
era\cite{IH-pt}, 
 (B-2) non-adiabatic evolution\cite{Modulated-matsuda} or
(B-3) the inhomogeneities in the fraction of the contents that have
different scaling (curvatons)\cite{curvaton-paper}. 
The case (B-1) includes inhomogeneous end of inflation\cite{End-Modulated,
End-multi-mat}, inhomogeneous reheating\cite{IH-R} or 
inhomogeneous phase transition\cite{IH-pt}. 
This paper considers the cases (B-1) and (B-2) for the inflationary
stage.
Although curvatons (B-3) or the creation of the curvature perturbations
 after inflation (A-4 and A-5) may lead to a significant result in the
generalized gravity theory,\footnote{See Ref.\cite{matsuda-gen-Pr} for
an example with scalar-tensor gravity.} 
they usually depend on the specific choice of the inflationary model and
the fields in the effective action. 
Therefore, for simplicity of the argument, we do not consider the
creation of curvature perturbations after inflation (A-4 and A-5),
which includes inhomogeneous reheating/preheating and the curvatons.

Considering the discussions of multi-field inflation, there are at least
      three ways to reach the same result of the cosmological
      perturbations. 
One is the straight calculation of $\dot{\cal R}$, which is considered
      in Ref.\cite{multi-orig-gau}.
The calculation is based on the field equations and the perturbations.
Another is to calculate the evolution of the ``perturbed expansion rate''
\cite{A-NEW, Modulated-matsuda} using the equations for the energy
      momentum and the perturbations, which is considered in this
      manuscript. 
The calculation based on this method is very easy and straight compared
      with the former one. 
The last method, which is intuitive but may be rather sketchy, is to
      calculate first the total number of e-foldings, then calculate 
$\delta N$.
This method must be complemented by the study of the evolution during
      inflationary stage\cite{warm-matsuda}.

\section{Curvature perturbations in generalized gravity theories}
In this paper, the Lagrangian that defines the generalized $f(\phi, R)$
gravity is given by
\begin{equation}
{\cal L}=\frac{1}{2}f(\phi^I, R)+P(X, \phi^I),
\end{equation}
where $X$ is defined by $ X\equiv -\frac{1}{2}G_{IJ}\partial_\mu
\phi^I\partial^{\mu}\phi^J$.
Here $\phi^I$ and $R$ are the scalar fields and the scalar curvature.
We consider the separation
\begin{equation}
P(X, \phi^I)=K(X, \phi^I)-V(\phi^I),
\end{equation}
and set $8\pi G=1$ for simplicity.
We define the energy-momentum in the standard way\cite{Hwang-Noh}.
General coordinate invariance suggests that the energy-momentum tensor
follows the conservation law, 
which is true without using the Einstein field equations.

It is possible to characterize cosmological perturbations using
the line element with linear scalar perturbations of a
Friedman-Lemaitre-Robertson-Walker(FLRW) background, which 
can be written as 
\begin{equation}
ds^2=-(1+2A)dt^2 +2 a^2(t)[\nabla_iB]dtdx^{i}
+a^2(t)[(1-2\psi)\gamma_{ij}+2\nabla_i\nabla_j E]dx^i dx^j.
\end{equation}
In FLRW spacetime, which is spatially homogeneous and isotropic,
from the variation of the metric $g_{\mu\nu}$ we get the gravitational
field equation and the energy momentum tensor.
Assuming a simple model of two-field inflation with a standard kinetic
term $K=X$, we obtain an energy density $\rho$ and pressure $p$
given by 
\cite{Hwang-Noh}
\begin{eqnarray}
\rho &=& \frac{1}{F}\left[X + \frac{RF-f}{2}+V -3H \dot{F}\right]
\nonumber\\
p&=& \frac{1}{F}\left[X - \frac{RF-f}{2}-V+2H\dot{F}+\ddot{F}\right],
\end{eqnarray}
where a dot ($\dot{\,\,}$) above a symbol 
denotes a derivative with respect to time.
In this paper we consider the adiabatic inflaton field $\phi$ and
the so-called entropy  field $s$ for two-field inflation.
Using the adiabatic inflaton field $\phi$, $X$ is redefined as 
$X\equiv \frac{1}{2}\dot{\phi}^2$ in FLRW spacetime.
$F$ is defined as $F\equiv f_{R}\equiv\partial f/\partial R$.
Here $\ddot{F}$ can not be neglected even in the slow-rolling
phase.\footnote{A useful example is given by $F=\phi^2$, which leads to
$\ddot{F}=2\dot{\phi}^2 + 2\ddot{\phi}\dot{\phi}\simeq 4X$ for
slow-rolling approximation $\ddot{\phi}\simeq 0$.}
We thus find
\begin{equation}
\rho+p=\frac{1}{F}\left[2X -H \dot{F}+\ddot{F}\right].
\end{equation}

Our interest is in the evolution of the curvature perturbation $\psi$
on the constant-time hypersurfaces of the FLRW background.
Introducing a normalized four vector field $n^\mu$ orthogonal
to the constant-time hypersurfaces, the expansion of the spatial
hypersurfaces with respect to the proper time $d\tau\equiv (1+A)dt$
is given by
\begin{equation}
\theta \equiv n^\mu_{;\mu}=3H(1-A)-3\dot{\psi}+\nabla^2 \sigma,
\end{equation}
where $\sigma$ is the scalar describing the shear, which is defined by
$\sigma\equiv \dot{E}-B$.
A useful definition of the expansion rate with respect to the
coordinate time is\cite{A-NEW}
\begin{equation}
\tilde{\theta}=(1+A)\theta =3H-3\dot{\psi}+\nabla^2\sigma,
\end{equation}
which can be used to define the perturbed expansion 
$\delta \tilde{\theta}\equiv \tilde{\theta}-3H$.
Without using the gravitational field equations, 
the energy conservation equation $n^\nu T^\mu_{\nu;\mu}=0$ for the
first-order density perturbations can be evaluated.
The equation is given by \cite{A-NEW}
\begin{equation}
\label{enemoeq}
\dot{\delta \rho}=-3H(\delta \rho+\delta p)
+(\rho+p)\left[3\dot{\psi}-\nabla^2 (\sigma + v + B)\right],
\end{equation}
which gives the equation of $\dot{\psi}$ in terms of the
local conservation of energy momentum.\footnote{The above equation
 is written for an ideal fluid, however
 in the generalized gravity, the fluid is in general not of the
 ideal form. This does not affect the first order analysis of the
 current work, since the non-ideal corrections to this equation are
 either of second order or contain spatial gradients which can be
 neglected on large scales. We appreciate a reviewer of the journal who
 sent us the argument related to the second-order perturbations in
 generalized gravity.}
The gauge-invariant combinations for the curvature
perturbation are usually defined by
\begin{eqnarray}
\label{zeta-org}
\zeta &=&-\psi -H\frac{\delta \rho}{\dot{\rho}}\nonumber\\
{\cal R}&=& \psi -H\frac{\delta q}{\rho+p},
\end{eqnarray}
where $\delta q$ is the momentum perturbation satisfying
\begin{equation}
\epsilon_m=\delta \rho-3H \delta q.
\end{equation}
Here $\epsilon_m$ is the perturbation of the comoving density,
satisfying the evolution equation 
\begin{equation}
\label{decay-comv}
\epsilon_m =-\frac{1}{4\pi G}\frac{k^2}{a^2}\Psi,
\end{equation}
where $\Psi$ is related to the shear perturbation, which is assumed to
be finite.
The continuity equation $\dot{\rho}=-3H(\rho+p)$, which can be used to
relate these definitions, does not depend on the gravitational equations.
The following useful expression can be straightforwardly obtained:
\begin{equation}
N \equiv \int_{t_{ini}}^{t_e} H dt
=\int_{\rho(t_{ini})}^{\rho(t_e)}H\frac{\delta \rho}{\dot{\rho}}.
\end{equation}
From the definition of $\zeta$, and using Eq.(\ref{enemoeq}) to
eliminate $\dot{\psi}$, a simple equation for the evolution of
$\zeta$ is given by
\begin{eqnarray}
\label{t-der-zeta2}
\dot{\zeta} &\simeq&
-\frac{\dot{\delta \rho}+3H(\delta \rho + \delta p)}{3(\rho+p)}
+\frac{d}{dt}\left[H\frac{\delta \rho}{3H(\rho+p)}\right]\nonumber\\
&=& -\frac{H}{\rho+p}
\left[\delta p-\frac{\dot{p}}{\dot{\rho}}\delta \rho \right]\nonumber\\
&=& -\frac{H}{\rho+p} \delta p_{nad},
\end{eqnarray}
where $\nabla^2 (\sigma + v + B)$, which should vanish at large
scales, is neglected in the first line.
To compare the above result with the $\delta N$
formalism\cite{delta-N-ini}, it is useful to define the $\delta N$
formalism in terms of the perturbed expansion rate $\delta
\tilde{\theta}$:
\begin{equation}
\zeta = \frac{1}{3}\int^t_{t_{ini}}\delta \tilde{\theta}dt =\delta N,
\end{equation}
where a hypersurface with a flat slicing at $t_{ini}$ and uniform
density at $t$ is chosen.
Therefore, the evolution equation for the $\delta N$ formalism is given
by 
\begin{eqnarray}
\label{del-n-t-2}
\frac{d}{dt}
\delta N &\equiv& \left.
\frac{1}{3}\delta \tilde\theta\right|_{\delta \rho=0}  \nonumber\\
&\simeq& -\left.\dot{\psi}\right|_{\delta \rho=0}\nonumber\\
&=& \left.\dot{\zeta} +
\frac{d}{dt}\left(H\frac{\delta \rho}{\dot{\rho}}\right)
\right|_{\delta \rho=0}
\nonumber\\
&=&-\left.\frac{H}{\rho+p} \delta p_{nad} +
\frac{d}{dt}\left(H\frac{\delta \rho}{\dot{\rho}}\right)\right|_{\delta \rho=0}.
\end{eqnarray}
Eqs. (\ref{t-der-zeta2}) and (\ref{del-n-t-2}) are
consistent as far as the $\delta N$ formalism is defined for a uniform
density slice at $t$. 
Our goal in this paper is to find the evolution equation for the
curvature
perturbation $\zeta$, which can be defined in terms of the $\delta N$
formalism;
\begin{equation}
\label{zetaN}
\dot{\zeta}_N\equiv -\left.\dot{\psi}\right|_{\delta \rho=0}
\simeq \left.-H
\frac{\delta \rho+\delta p}{\rho+p}\right|_{\delta \rho=0},
\end{equation}
where $\dot{\psi}$ is obtained from Eq.(\ref{enemoeq}) 
on uniform density hypersurfaces ($\dot{\delta \rho}=0$).
$\delta \rho$ in the last equation is for the later convenience. 
In usual inflation for the standard-gravity theory, 
$\dot{\zeta}_N$ is related to the perturbation of the adiabatic
inflaton velocity using the relation 
$\delta(\rho+P)\sim \delta (\dot{\phi}^2)$ \cite{Modulated-matsuda}.
This result is convincing as the origin of the inhomogeneities of
the time elapsed during inflation ($\delta t$) is generally given by the
inhomogeneities of the ``length'' ($\delta \phi$) and the ``velocity''
($\dot{\phi}$).\footnote{The perturbation of the kinetic term sources
the creation of the curvature perturbations at the ``bend'' in the
trajectory of multi-field inflation.\cite{Modulated-matsuda,
warm-matsuda}. After submitting this paper to the journal, we find
papers\cite{add-after-sub} in which entropy perturbations are considered
for generalized gravity.} 
This simple result is not true for generalized gravity theories,
as we will show in the following.
An intuitive argument would be useful for the argument.
In the generalized gravity theory, the gravitational constant may 
not be a homogeneous constant, which may source the inhomogeneities of
the expansion rate.
The situation is similar to the scalar-tensor gravity theory considered
in Ref.\cite{multi-orig-gau, Hwang-Noh, Modulated-matsuda}.

The perturbations of the energy density and the pressure are given
by\cite{Hwang-Noh} 
\begin{eqnarray}
\delta \rho &\simeq
& \frac{1}{F}\left[\delta X
+\frac{F\delta R  + R \delta F -f_R\delta R- f_\phi \delta \phi
-f_s\delta s}{2}
+V_\phi \delta \phi +V_s\delta s \right.\nonumber\\
&& \,\,\,\, \left.-3H\delta [\dot{F}]- \dot{F}\delta\theta\right]
-\frac{\delta F}{F}\rho\\
&=& \frac{1}{F}\left[\delta X
+\frac{(R-2\rho) \delta F- f_\phi \delta \phi
-f_s\delta s}{2}\right.\nonumber\\
&&\,\,\,\, \left.+V_\phi \delta \phi+V_s\delta s
 -3H\delta [\dot{F}]- \dot{F}\delta\theta\right]\\
\delta p&\simeq& 
\frac{1}{F}\left[\delta X - \frac{(R+2p)\delta F-f_\phi\delta \phi
-f_s\delta s}{2}\right.\nonumber\\
&&\left.\,\,\,\, -V_\phi\delta \phi-V_s\delta s
  +2H\delta[\dot{F}]+\frac{2}{3}\dot{F}\delta \theta+\delta[\ddot{F}]\right],
\end{eqnarray}
where $\delta X\equiv \dot{\phi}(\delta \dot{\phi}-\dot{\phi}A)$
and $\delta [\dot{F}]\equiv \delta \dot{F} -\dot{F}A$ are introduced.
Here $\delta \theta$ denotes the perturbation of the expansion scalar
$\theta$, which follows the perturbed equation given
by\cite{Hwang-Noh}
\begin{equation}
H\delta \theta=-\frac{1}{2}\delta \rho + \frac{k^2}{a^2}\psi,
\end{equation}
which suggests that $\delta \theta$ is negligible at large scales 
on uniform density hypersurfaces. 
From the above equations we find
\begin{equation}
\delta (\rho+p)=
\frac{1}{F}\left[2\delta X - (\rho+p)\delta F
-H\delta[\dot{F}]-\frac{1}{3}\dot{F}\delta \theta+\delta[\ddot{F}] \right].
\end{equation}
For the momentum perturbation it is found that\cite{Hwang-Noh}
\begin{equation}
\delta q = \frac{1}{F}\left[
-\dot{\phi}\delta \phi-\delta [\dot{F}]+H\delta F\right].
\end{equation}

Considering the evolution of $\epsilon_m$ in the {\bf standard} gravity
theory, where $F=1$ is assumed, it is found that $\delta q$
vanishes at large scales on uniform density hypersurfaces.
In terms of the $\delta N$  formalism, $\delta q\simeq 0$
leads to a natural condition $\phi\simeq$const. at large scales, which
can be interpreted as a flat boundary $(\delta \phi_e =0)$ 
at the end of inflation appearing in the standard inflationary scenario.
On the other hand, in {\bf generalized} gravity theories the uniform
density hypersurfaces does not always lead to a flat ($\delta \phi=0$)
boundary at $t$. 
This result shows that in contrast to the standard gravity theory the
generation of the curvature perturbation at 
the end of inflation, which is usually evaluated using the $\delta N$
formalism\cite{End-Modulated}, is not trivial in generalized gravity
theories.\footnote{See also the discussions in Ref.\cite{Hwang-Noh}.} 
Namely, considering the curvature perturbation defined by ${\cal R}$,
the $\delta N$ formula can lead
 to the boundary perturbations $\delta N_{ini}$ and $\delta
N_e$ defined by
\begin{eqnarray}
\delta N_{ini}&\equiv& H\frac{\delta q_{ini}}{\rho+p}\nonumber\\
\delta N_{e}&\equiv& -H\frac{\delta q_{e}}{\rho+p}.
\end{eqnarray}
In standard gravity theory, the perturbation at the end-boundary
($\delta N_e$)  
vanishes if $\delta q_e =0$, which directly leads to the flat boundary
condition $\phi_e$=const.
However, in generalized gravity theories, the flat boundary at 
$\phi=\phi_e$ does not always lead to $\delta q_e=0$, but rather to the
perturbation $\delta q_{e}\simeq \frac{1}{F}\left[-\delta [\dot{F}]
+H\delta F\right]$, which leads to 
\begin{eqnarray}
\delta N_{e}&=& -\frac{H}{\rho+p}\frac{1}{F}\left[-\delta [\dot{F}]
+H\delta F\right]_{\delta \phi=0}\ne 0.
\end{eqnarray}
Here $\delta N_e$ is defined for the expansion between
a hypersurface with a uniform density slicing ($\delta q=0$) and 
uniform field ($\delta \phi=0$) at $t=t_e$.
The above correction does not appear in the standard gravity theory.
The perturbation $\delta N_{ini}$ on the initial boundary is given by
\begin{eqnarray}
\delta N_{ini}&=& \frac{H}{\rho+p}\frac{1}{F}\left[
-\dot{\phi}\delta \phi-\delta [\dot{F}]
+H\delta F\right]_{\psi=0}.
\end{eqnarray}
For two-field inflation at first-order expansion, $\delta F$ is given by
$\delta F \simeq F_\phi\delta \phi + F_s \delta s$,
where $\phi$ and $s$ denote the adiabatic component of the multi-field
inflaton and the entropy field.
On the uniform-field($\delta \phi=0$) hypersurfaces, the expansion is
given by 
$\delta F \simeq F_s \delta s$.
Assuming that in the slow-rolling phase, the rate of variation of
cosmological quantities is much smaller than unity, the sum of the
boundary perturbations leads to\footnote{This
simple assumption is not true if there is a sharp bend in the
trajectory.}
\begin{eqnarray}
\delta N_{ini}+\delta N_e &\simeq& \frac{H}{\rho+p}\frac{1}{F}\left[
-\dot{\phi} -\frac{\partial [\dot{F}]}{\partial \phi}
+H F_\phi \right]\delta \phi,
\end{eqnarray}
where only the perturbations caused by the adiabatic field $\phi$ can
survive  because of the uniform field condition at the end ($t=t_e$)
boundary. 

Creation of perturbations at the boundaries (and
during inflation) depends on the definition of the hypersurfaces for the
$\delta N$ formalism,
while the sum for the inflationary period is independent.
$\delta N_e\ne0$ appears because the standard definition of the
$\delta N$ formalism, which has been used in this paper,
 is for flat hypersurface at $t_{ini}$ and uniform
density at $t$.
In contrast to the above argument, which is for boundary
perturbations related to the $\delta N$ formalism defined by the uniform
density hypersurfaces at $t$, it would be possible to define the $\delta
N$ formalism for  flat hypersurfaces at $t_{ini}$
and a uniform ``field'' ($\phi=$const.) at $t$.
In the latter case, no perturbations are generated at the end of
inflation (except for a specific case in which the inhomogeneous end of
inflation is sourced by the entropy field).

To show the equivalence between the $\delta N$
formalism defined by different hypersurfaces, it would be useful
(although it might be trivial for an expert) to
 consider an explicit calculation of the time integrel. 
Considering a separation
\begin{eqnarray}
\frac{d}{dt}
\delta N 
&=&-\frac{H}{\rho+p} \delta p_{nad} +
\frac{d}{dt}\left(H\frac{\delta \rho}{\dot{\rho}}\right),
\end{eqnarray}
the time-integral of the last term vanishes if the $\delta N$ formalism
is defined for uniform density hypersurfaces.
However, it does not vanish for uniform ``field'' hypersurfaces
and gives
\begin{eqnarray}
\int_{t_{ini}}^{t_e}
\left[\frac{d}{dt}\left(H\frac{\delta \rho}{\dot{\rho}}\right)\right]dt
&=&\left(H\frac{\delta \rho}{\dot{\rho}}\right)_{t=t_e}-
\left(H\frac{\delta \rho}{\dot{\rho}}\right)_{t=t_{ini}}\nonumber\\
&\simeq&\left(H\frac{\delta q|_{\delta \phi=0}}{\rho+p}\right)_{t=t_e}-
\left(H\frac{\delta q|_{\delta \phi=0}}{\rho+p}\right)_{t=t_{ini}},
\end{eqnarray}
where $\delta q$ is defined for uniform ``field'' hypersurfaces.
The sign of $\delta N$ defined for the expansion from uniform density
hypersurfaces to uniform ``field'' hypersurfaces is opposite to the one 
defined for the expansion from uniform ``field'' hypersurfaces to
uniform density hypersurfaces.
Obviously, the sum of the perturbations $\delta N \equiv \delta
N_{ini}+\delta N_e + \int \dot{N}dt$ gives the
identical result for different choices of the hypersurfaces.
Equations for the boundary perturbations and the
evolution in terms of the $\delta N$ formalism depend on the
definition of the hypersurfaces, while the sum of these
quantities gives identical result.

The explicit form of the evolution equation can be evaluated with the
help of field equations.
For two-field inflation with the standard kinetic term $K=X$ and
$G_{IJ}=\delta_{IJ}$, the equation of motion for the adiabatic inflaton
field is given by
\begin{equation}
\ddot{\phi}+3H\dot{\phi}-\frac{f_\phi}{2}+V_\phi=0.
\end{equation}
From the trace of the gravitational field equation, it is found
that\cite{Hwang-Noh} 
\begin{eqnarray}
\label{trace-gra}
\ddot{F}&+&3H \dot{F}+\frac{1}{3}\left[2X-RF+2f-4V
\right]=0\\
R&=&\rho -3p.
\end{eqnarray}
As $\epsilon_m$ decays at large scales as $\propto k^2/a^2$,
the comoving energy density on uniform density hypersurfaces leads to,
\begin{equation}
\epsilon_m \simeq \frac{1}{F}\left[\delta X -\frac{3}{2}(\rho+p)\delta F 
-\frac{f_s}{2}\delta s +V_s\delta s
-\ddot{\phi}\delta \phi\right]\simeq 0,
\end{equation}
which gives the equation for $\delta X$:
\begin{equation}
\label{dx-org}
\delta X \simeq \frac{3}{2}(\rho+p)\delta F 
+\frac{f_s}{2}\delta s -V_s\delta s
+\ddot{\phi}\delta \phi.
\end{equation}
At large scales on uniform density hypersurfaces, we thus find
\begin{eqnarray}
\label{delrodelp}
\delta \rho+\delta p &\simeq & 
\frac{1}{F}\left[2\delta X -(\rho+p)\delta F -H\delta [\dot{F}]
+\delta[\ddot{F}]\right]
\nonumber\\
&\simeq & 
\frac{1}{F}\left[2(\rho+p)\delta F 
+f_s\delta s -2V_s\delta s  -H\delta [\dot{F}]+\delta[\ddot{F}]\right].
\end{eqnarray}
A useful expression for the evolution of the curvature perturbation 
obtained from the above equation is given by
\begin{equation}
\dot{\zeta}_N \simeq -2H\frac{\delta F}{F}
-\frac{H}{F(\rho+p)}\left[(f_s-2V_s)\delta s -H\delta [\dot{F}]
+\delta[\ddot{F}]
\right].
\end{equation}
Here $\delta F$ must be evaluated on uniform density hypersurfaces,
which (in contrast to the standard gravity theory) does not lead to 
$\delta \phi =0$ hypersurfaces, as we discussed for the boundary
perturbations. 
For the standard gravitational theory with $F=1$, this equation 
gives the well-known result\cite{2-field}
\begin{equation}
\dot{\zeta}_N \simeq 2H\frac{V_s}{\dot{\phi}^2}\delta s,
\end{equation}
which agrees with the evolution of the curvature perturbation 
during inflation in standard gravity theory.

The perturbed equation from Eq.(\ref{trace-gra}) is given by
\begin{eqnarray}
\label{dotf}
\delta[\ddot{F}]+ 3\dot{F}\delta \theta +3H \delta[\dot{F}]
&\simeq&-\frac{1}{3}\left[2\delta X-R\delta F+F\delta R-4\delta V\right],
\end{eqnarray}
which shows that $\delta[\ddot{F}]$ can be expressed by using 
$\delta[\dot{F}]$ and other quantities. 
Here a reasonable approximation would be $F_R \dot{R}\ll F_\phi \dot{\phi}$,
which makes it possible to calculate $\delta[\dot{F}]$ in terms of 
Eq.(\ref{dx-org}) for specific models of $f(\phi,R)$ gravity.

\section{Conclusion and discussion}
In this paper the curvature perturbations from the boundaries and the
evolution of the curvature perturbations are
considered for generalized gravity theories in terms of the $\delta N$
formalism.

Our result shows that the creation of the curvature perturbation may be
significant during the evolution.
The perturbation $\delta F$ can be interpreted as a spatial
inhomogeneities in the gravitational constant\cite{Modulated-matsuda}.
The second-order perturbations with respect to the field
perturbations can be used to estimate the order of the magnitude of 
the non-Gaussian character of the cosmological
perturbations.\footnote{Our equations are not exact in the second-order
expansion. They can be used to estimate the order of magnitude of
the non-Gaussian character, only when there is no significant source
other than the trivial field perturbations.} 

In this paper, we considered the generalized gravity that is
expressed by $f(\phi^I,R)$.
This model is suitable for our present study, as it gives a very clear 
and intuitive result.
However, it would be useful to make some comments for a typical 
string corrections that may not appear as $f(\phi^I,R)$.
For example, the effective action may appear with the following
additional corrections in the action\cite{Hwang-Noh}
\begin{equation}
{\cal L}_{add}=\frac{1}{2}\xi(\phi) R^2_{GB},
\end{equation}
where $R^2_{GB}\equiv R^{abcd}R_{abcd}-4R^{ab}R_{ab}+R^2$.
Using the explicit form of the energy momentum tensor given in 
Ref.\cite{Hwang-Noh}, it is indeed possible to show the explicit 
formula of the perturbed expansion rate after a lengthy
calculation.\footnote{The evolution of the curvature perturbations in
terms of the perturbed expansion rate and the $\delta N$ formalism 
(the formulation used in this paper) is discussed in
Ref.\cite{matsuda-toappear} for the effective action with generalized
kinetic terms. DBI action in string theory is a typical example of this
kind\cite{add-after-sub}.}

Moreover, recent study shows that ultraviolet-complete quantum theory of
 gravity (Ho\v rava-Lifshitz Gravity\cite{horava-gra})
 may be given by a (more) generalized gravity action.
In the generalized gravity theory considered in this paper, a 
crucial difference appears in the end-boundary of the inflationary
 stage, which is due to the non-ideal 
 form of the energy momentum tensor that depends explicitly on the
 curvature scalar.
In this sense, our paper may give an important step in understanding the
 discrepancy between 
the ultraviolet-complete theory and the conventional gravity in 
terms of the evolution of the curvature perturbation during inflation,
 where the energy momentum tensor may not be given by the ideal form
due to the ultraviolet corrections from the fundamental theory.

\section{Acknowledgment}
We wish to thank K.Shima for encouragement, and our colleagues at
Tokyo University for their kind hospitality.

\end{document}